\documentclass[aip,jcp,reprint]{revtex4-2}

\usepackage{amssymb, amsmath} 
\usepackage{amsfonts}
\usepackage{physics}
\usepackage{bm}
\usepackage[utf8]{inputenc}
\usepackage[T1]{fontenc}
\usepackage{graphicx}
\usepackage{xcolor}
\usepackage{graphics}
\usepackage{newtxtext,newtxmath}
\usepackage{microtype}

\definecolor{cset-aps-blueberry}{RGB}{28,128,158}
\definecolor{cset-aps-blue}{RGB}{46,44,184}
\definecolor{cset-aps-turquoise}{RGB}{0,67,88}
\definecolor{cset-aps-limegreen}{RGB}{190,219,67}
\definecolor{cset-aps-green}{RGB}{31,138,112}
\definecolor{cset-aps-yellow}{RGB}{255,225,25}
\definecolor{cset-aps-orange}{RGB}{253,116,0}
\definecolor{cset-aps-red}{RGB}{219,0,43}

\usepackage[]{hyperref}
\hypersetup{%
    colorlinks=true,
    linkcolor={cset-aps-red},
    linkbordercolor={cset-aps-red},
    filecolor={cset-aps-orange},
    filebordercolor={cset-aps-orange},
    citecolor={cset-aps-blue},
    citebordercolor={cset-aps-blue},
    urlcolor={cset-aps-green},
    urlbordercolor={cset-aps-green},
    menucolor={cset-aps-limegreen},
    menubordercolor={cset-aps-limegreen},
    breaklinks=true,
    pdfborderstyle={/S/U/W 2},
    pdfpagemode=UseOutlines,
    pdfstartpage={1},
}

\newcommand{\ee}{\text{e}}
\newcommand{\ii}{\text{i}}
\definecolor{cset-aps-red}{RGB}{219,0,43}

\begin{document}


\title{Atom interferometry with quantized light pulses}
\collaboration{This article is part of the \emph{JCP Special Topic on Quantum Light} and has been published in\\
\href{https://doi.org/10.1063/5.0048806}{The Journal of Chemical Physics \textbf{154}, 164310 (2021)}; 
licensed under a \href{http://creativecommons.org/licenses/by/4.0/}{Creative Commons Attribution [CC BY]} license.\vspace{1em}}


\newcommand{\orcid}[1]{\href{https://orcid.org/#1}{\includegraphics[width=7pt,height=7pt]{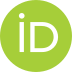}}}
\author{Katharina Soukup}
\author{Fabio Di Pumpo\,\orcid{0000-0002-6304-6183}}
\email[]{fabio.di-pumpo@uni-ulm.de}
\email[]{fabio.di-pumpo@gmx.de}
\author{Tobias Asano\,\orcid{0000-0002-6257-8815}}
\affiliation{Institut f{\"u}r Quantenphysik and Center for Integrated Quantum Science and Technology (IQ\textsuperscript{ST}), Universit{\"a}t Ulm, Albert-Einstein-Allee 11, D-89069 Ulm, Germany}
\author{Wolfgang P. Schleich\,\orcid{0000-0002-9693-8882}}
\affiliation{Institut f{\"u}r Quantenphysik and Center for Integrated Quantum Science and Technology (IQ\textsuperscript{ST}), Universit{\"a}t Ulm, Albert-Einstein-Allee 11, D-89069 Ulm, Germany}
\affiliation{Institute of Quantum Technologies, German Aerospace Center (DLR), S\"{o}flinger Stra\ss e 100, D-89077 Ulm, Germany}
\affiliation{Hagler Institute for Advanced Study and Department of Physics and Astronomy, Institute for Quantum Science and Engineering (IQSE), Texas A{\&}M AgriLife Research, Texas A{\&}M University, College Station, Texas 77843-4242, USA}
\author{Enno Giese\,\orcid{0000-0002-1126-6352}\,}
\affiliation{Institut f{\"u}r Quantenphysik and Center for Integrated Quantum Science and Technology (IQ\textsuperscript{ST}), Universit{\"a}t Ulm, Albert-Einstein-Allee 11, D-89069 Ulm, Germany}
\affiliation{Institut f{\"u}r Quantenoptik, Leibniz Universit{\"a}t Hannover, Welfengarten 1, D-30167 Hannover, Germany}
\affiliation{Current address: Institut für Angewandte Physik, Technische Universität Darmstadt, Schlossgartenstr. 7, D-64289 Darmstadt, Germany}



\begin{abstract}
The far-field patterns of atoms diffracted from a classical light field, or from a quantum one in a photon-number state are identical.
On the other hand, diffraction from a field in a coherent state, which shares many properties with classical light, displays a completely different behavior.
We show that in contrast to the diffraction patterns, the interference signal of an atom interferometer with light-pulse beam splitters and mirrors in intense coherent states does approach the limit of classical fields.
However, low photon numbers reveal the granular structure of light, leading to a reduced visibility since Welcher-Weg (which-way) information is encoded into the field.
We discuss this effect for a single photon-number state as well as a superposition of two such states.
\end{abstract}


\maketitle

\section{Introduction}
During the last decades, the interaction of atoms with quantized light fields~\cite{Freyberger1999} has led to landmark experimental achievements, such as the one-atom maser~\cite{Meschede1985,Varcoe2000} or the generation of Schrödinger-cat states~\cite{Brune1996}.
At the same time, light-pulse atom interferometers~\cite{Kasevich1991} have become unique instruments for precision measurements.
In this article we, analyze the interference signal of a Mach-Zehnder atom interferometer where we have replaced the classical light creating the beam splitters and mirrors by quantum fields.
Our analysis also shines some light on a surprising fact~\cite{Akulin1991}:
The underlying mechanism of diffraction of an atom from a standing wave leads to identical momentum distributions for classical light and a photon-number state.
However, the most classical state, that is a coherent state, causes a momentum distribution which is utterly different.
We study whether such a behavior also transfers to the interference signal observed in atom interferometers~\cite{Bongs2019} generated from light pulses~\cite{Soukup2020}.

\subsection{Classical versus quantum field}
\begin{figure}
	\centering
	\includegraphics[width=1\columnwidth]{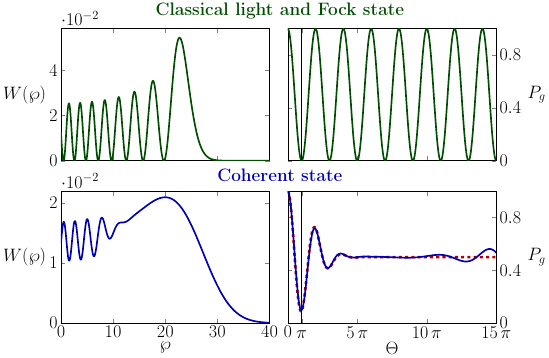}
	\caption{
	Classical versus quantum.
	A classical electromagnetic field and a field in the most non-classical state, that is a Fock state, cause identical far-field diffraction patterns and identical Rabi oscillations.
	However, the corresponding curves for the most classical state, that is a coherent state, are substantially different.
	The left side shows the momentum distribution $W(\wp)$ of an atom after diffraction from a standing wave as a function of the dimensionless momentum $\wp \equiv p/(\hbar k)$, for the pulse area of $\Theta=8\pi$.
	For classical light as well as for Fock states (top) the distribution is given by a Bessel function with oscillations of perfect contrast.
	However, for a coherent state (bottom) with mean photon number $\bar{n}=|\alpha|^2=6$, the interference pattern is washed out due to the average over the photon distribution.
    The right side shows the probability $P_g$ of the undiffracted momentum for atomic Bragg or Raman diffraction as a function of $\Theta$.
	For classical light and Fock states (top) we observe perfect Rabi oscillations.
	Conversely, for a coherent state with $\bar{n}=|\alpha|^2=6$ a collapse and a revival emerge caused by the average.
	For sufficiently large $|\alpha|^2$, the collapse and initial dephasing is well described by the approximation of Eq.~\eqref{eq:ProbCohStateApprox} (red dashed line).
	In such a case or for small pulse areas, $P_g$ approaches the Rabi limit, where dephasing effects are negligible.
	The vertical line denotes the pulse area $\Theta= \pi$ used for mirror pulses.}
    \label{fig:Probabilities}
\end{figure}
Atom interferometry~\cite{Kasevich1991} offers a powerful tool for a wealth of applications such as gravimetry, inertial sensing, and metrology~\cite{Abend2016} but also for fundamental tests of physics such as the Einstein equivalence principle~\cite{Schlippert2014,Ufrecht2020}.
These schemes usually rely on the diffraction from intense classical light pulses~\cite{Rasel1995}.

However, optical cavities represent a promising route to enhance~\cite{Hamilton2015,Canuel2018} the sensitivity of an atom interferometer.
Indeed, configurations based on cavities offer major experimental advantages such as higher intensities, a reduction of wave-front distortions, and clearer mode profiles. 
Moreover, diffraction in cavities can be employed for the generation of entanglement~\cite{Khosa2004,Haine2016} or squeezed atomic states~\cite{Salvi2018}, and for quantum non-demolition measurements~\cite{Bernon2011,Shankar2019}.
In light of such developments, it seems natural to study the influence of quantized light on atom interferometry, even though these implementations are yet far away from the quantized regime.

Atomic diffraction from a classical standing wave in the Raman-Nath regime~\cite{Domokos1996,Rohwedder2000} leads to the momentum distribution~\cite{Akulin1991,Herkommer1992,Rohwedder2000}
\begin{equation}
\label{eq:Bessel}
    W(\wp)=J^2_{\wp}(\Theta), 
\end{equation}
where $J_s(\kappa)$ is the Bessel function of first kind of order $s$ and argument $\kappa$, and $\wp \equiv p /(\hbar k)$ denotes the momentum in units of the momentum transfer $\hbar k$ with the wave vector $k$ of the light field.
Here, $\Theta$ describes the pulse area determined by the intensity of the classical field and the  duration of the interaction.

Surprisingly, also for a photon-number state, that is a Fock state, even though highly non-classical, one finds \emph{exactly} the same diffraction pattern, where the intensity is given by the photon number~\cite{Akulin1991}.

In contrast, for a coherent state, which shares many properties with classical light, the momentum distribution
\begin{equation}
    W(\wp)=\sum\limits_{n=0}^{\infty}{W_n J^2_{\wp}\left(\Theta \sqrt{\frac{n}{\bar{n}}} \right)}
\end{equation}
involves an average of Eq.~\eqref{eq:Bessel} over the Poissonian photon distribution
\begin{equation}
\label{eq:coherent_state_distribution}
    W_n \equiv \frac{|\alpha|^{2n}}{n!}\ee^{-|\alpha|^2}
\end{equation}
of the coherent state with $|\alpha|^2=\bar{n}$. 
Figure~\ref{fig:Probabilities} shows that the interference pattern of the Bessel function is washed out and therefore displays a behavior which is completely different from that of classical fields and Fock states.

We now transfer these observations to the Bragg regime where higher diffraction orders are suppressed~\cite{Giltner1995,Giese2013,Giese2015}.
As a consequence, the field drives Rabi oscillations between momentum states of the atom, encoded in a transition probability $P_g$.
Similar to the diffraction pattern in the Raman-Nath regime, the transition probabilities $P_g$ induced by classical light fields as well as Fock states behave alike~\cite{Akulin1991}, i.\,e. we observe the oscillations depicted on the top right of Fig.~\ref{fig:Probabilities}.
However, as can be seen from the bottom right of the figure, a coherent state leads to a collapse and revival of the oscillations~\cite{Cummings1965,Eberly1980,Knight1982,Fleischhauer1993}.

These observations have direct consequences for the atom-optical elements in light-pulse atom interferometry, since beam splitters and mirrors are usually performed in the Bragg regime.
To analyze such effects, we utilize a model based on Raman diffraction~\cite{Kasevich1991} with running light waves instead of standing ones.

\subsection{Overview and outline}
In this article we demonstrate that (i)~the phase of a coherent state contributes to the phase of a Mach-Zehnder interferometer in the same way as the phase of classical light.
(ii)~The visibility of the interference signal generated by diffraction from coherent states approaches the limit of diffraction from classical light pulses for high average photon numbers~\cite{Bertet2001}.
(iii)~However, for diffraction from a Fock state the visibility vanishes~\cite{Agarwal2003}, since complete Welcher-Weg~\cite{Scully1991,Storey1994,Englert1995,Rasel1995,Duerr1998} information can be inferred.
(iv)~Even for a superposition of two Fock states in every light field, we observe a significant loss of visibility. 
In addition, the interferometric phase differs from the one obtained by diffraction from classical fields.

Our article is organized as follows: In Sec.~\ref{sec.Atomic_diffraction} we summarize the essential ingredients of atomic diffraction from quantized light fields.
In particular, we establish the corresponding scattering operator which plays a central role throughout our analysis, and compare and contrast the atomic diffraction as well as the Rabi oscillations due to a coherent state and a Fock state.
We then turn in Sec.~\ref{sec.Mach-Zehnder_atom_interferometer} to our description of a Mach-Zehnder interferometer where the beam splitters as well as the mirrors are either classical or in an arbitrary quantum state.
With the help of the scattering operator we express the unitary time evolution along the two paths of the interferometer by two operator sequences which differ in their order, and derive an expression for the interference signal.
In Sec.~\ref{sec.Examples_of_quantum_light} we illustrate the results of Sec.~\ref{sec.Mach-Zehnder_atom_interferometer} for the cases of a coherent state, a single Fock state, and a superposition of two Fock states.
We conclude in Sec.~\ref{sec.Conclusions} by summarizing our main results and by providing an outlook.

\section{Atomic diffraction}
\label{sec.Atomic_diffraction}
The atom-optical analogs to beam splitters and mirrors are light pulses that diffract the atom and generate superpositions of different momenta.
In this section, we present an elementary extension~\cite{Schleich2001} of the Jaynes-Cummings model~\cite{Cummings1965}, taking into account the center-of-mass (COM) motion of the atom. 
We then analyze the effect of a photon distribution on the Rabi oscillations.

\subsection{Scattering operator}
Even though atomic Raman diffraction is a two-photon process~\cite{Schleich2013,Hartman2020,Hartman2020_2} stimulated by two counterpropagating light fields, it can be modeled by an effective two-level system~\cite{Schleich2001,Schleich2013} coupled to a single running wave~\cite{Riehle1991}.
In contrast to the conventional semiclassical treatment, we consider in this article a quantized light field.

For monochromatic plane-wave fields and short pulses we describe~\cite{Haine2013,Kleinert2015} the resonant diffraction process from a single mode by the effective scattering operator
\begin{align}
\label{eq:ScattMatrix}
\begin{split}
    \hat{\mathcal{S}} &\equiv  c_{\hat{n}+1} \, \op{e} -\ii \, \ee^{\ii \, \qty( k \, \hat{z} +\theta)} \, \hat{a} \, \frac{s_{\hat{n}}}{\sqrt{\hat{n}}} \, \op{e}{g}  \\
    &-\ii \, \ee^{-\ii \, \qty(  k \,\hat{z} +\theta)} \, \frac{s_{\hat{n}}}{\sqrt{\hat{n}}} \, \hat{a}^{\dagger} \, \op{g}{e} + c_{\hat{n}} \, \op{g}.
    \end{split}
\end{align}
Here, the photonic annihilation and creation operators $\hat{a}$ and $\hat{a}^{\dagger}$ obey the conventional bosonic commutation relation
\begin{equation}
    \big[\hat{a},\hat{a}^{\dagger}\big]=1.
\end{equation} 

The associated momentum transfer $\pm\hbar k$ is represented by the displacement operator $\exp(\pm \ii k \hat{z})$, where $\hat{z}$ is the COM position of the atom.
In general, the coupling constant between the atom and the light field is complex, leading to a phase $\theta$.

Additionally, we have introduced the abbreviations
\begin{subequations}
\begin{equation}
    c_{\hat{n}} \equiv\cos\left(\frac{\Theta}{2}\sqrt{\frac{\hat{n}}{\Bar{n}}}\right)
\end{equation}
and
\begin{equation}
    s_{\hat{n}} \equiv\sin\left(\frac{\Theta}{2}\sqrt{\frac{\hat{n}}{\Bar{n}}}\right),
\end{equation}
\end{subequations}
where $\Theta$ is the pulse area defined by the mean photon number $\bar n$, and $\hat{n}\equiv \hat{a}^{\dagger}\hat{a}$ represents the number operator of the field. 

The scattering operator $\hat{\mathcal{S}}$ defined by Eq.~\eqref{eq:ScattMatrix} exhibits a rather intuitive structure.
We first note that it involves three quantum degrees:
(i) The two internal energy levels $\ket{g}$ and $\ket{e}$ of the atom, (ii) the electromagnetic field characterized by the annihilation and creation operators $\hat{a}$ and $\hat{a}^\dagger$, and
(iii) the COM motion of the atom given by the position operator $\hat{z}$, and entering into $\hat{\mathcal{S}}$ as the displacement operator in momentum.

Moreover, we note that $\hat{\mathcal{S}}$ consists of two distinct parts describing two types of unitary evolution:
(i) The atom remains in its original state $\ket{g}$ or $\ket{e}$ represented by the projection operators $\ket{g}\!\bra{g}$ and $\ket{e}\!\bra{e}$.
In this case, the trigonometric operators $c_{\hat{n}}$ and $c_{\hat{n}+1}$ associated with the Rabi oscillations appear, but no creation or annihilation operator changes the overall photon number.
Moreover, the COM motion remains unaffected due to the absence of the displacement operator.

(ii) The atom makes a transition from the ground to the excited state given by $\ket{e}\!\bra{g}$, or from the excited to the ground state corresponding to $\ket{g}\!\bra{e}$.
In both cases the transition is due to Rabi oscillations and involves the trigonometric operators $s_{\hat{n}}/\sqrt{\hat{n}}$.
However, going into the excited state requires the \emph{absorption} of a photon as expressed by the appearance of the annihilation operator $\hat{a}$, whereas the transition to the ground state is associated with \emph{emission} and thus with the creation of a photon as reflected by $\hat{a}^\dagger$.

Due to momentum conservation, these processes lead to a momentum exchange $\pm \hbar k$ between the photon and the COM of the atom, where $k$ is the wave vector of the photon.
Hence, the light-matter interaction influences the atomic trajectory and allows for the creation of spatial superpositions.
Indeed, the displacement operators originating from the plane electromagnetic wave decrease or increase the momentum of the atom by $\hbar k$ due to the elementary relation
\begin{equation}
    \ee^{\pm \ii k\hat{z}}\ket{p}=\ket{p \pm \hbar k}
\end{equation}
where $\ket{p}$ denotes an eigenstate of the momentum operator.

It is interesting to note that the prefactor $(-\ii)$ in front of the transition terms is a consequence of the imaginary unit appearing in the Schr\"odinger equation.
Indeed, because $\hat{\mathcal{S}}$ constitutes a solution of the Schr\"odinger equation, it is a unitary operator.
This unitarity is also the reason why the order of the two operators $s_{\hat{n}}/\sqrt{\hat{n}}$ and $\hat{a}$ or $\hat{a}^\dagger$ is reversed on the off-diagonal. 

\subsection{Rabi oscillations}
The effective scattering operator $\hat{\mathcal{S}}$ given by Eq.~\eqref{eq:ScattMatrix} represents the unitary time evolution of a given quantum state and allows us to evaluate quantum mechanical expectation values.
For example, for an atom initially in the ground state the quantity
\begin{equation}
    P_g=\langle c_{\hat{n}}^2\rangle
\end{equation} describes the ground state population.

For a classical field, we find the familiar Rabi oscillations 
\begin{equation}
    P_g=\cos^2( \Theta/2)
\end{equation}
shown on the top right of Fig.~\ref{fig:Probabilities}.

If the light field is initially in a highly non-classical Fock state~\cite{Varcoe2000}, we can simply replace $\hat{n}$ by $\bar{n}$ and find \emph{exactly} the same probability.

However, for a coherent state $\ket{\alpha}$, which is a good approximation of laser light, the probability 
\begin{equation}
    \label{eq.average_coh}
    P_g=\bra{\alpha} c_{\hat{n}}^2\ket{\alpha} = \sum\limits_{n=0}^\infty W_n \cos^2\left(\frac{\Theta}{2}\sqrt{\frac{n}{\bar n}} \right)
\end{equation}
involves an average over the photon distribution from Eq.~\eqref{eq:coherent_state_distribution}.
The scattering operator in general entangles the light field and the atom.
However, the average in Eq.~\eqref{eq.average_coh} is a consequence of the trace over the photonic system, which in principle leads to decoherence.

The result of such a procedure is depicted on the bottom right of Fig.~\ref{fig:Probabilities}, where this average leads to a dephasing and finally to a revival of the transition probability~\cite{Cummings1965,Eberly1980,Knight1982,Fleischhauer1993,Meunier2005}.
Obviously, this behavior differs significantly from the one caused by a Fock state or a classical field.

In order to gain some insight into the collapse we replace~\cite{Cummings1965,Eberly1980} in the limit of $|\alpha|^2>>1$ the Poissonian photon distribution of the coherent state by a Gaussian, and the discrete sum by an integral. 
Moreover, we expand the square root
\begin{equation}
   \sqrt{1+ \frac{n-|\alpha|^2}{|\alpha|^2}}\cong1+\frac{n-|\alpha|^2}{2|\alpha|^2}
\end{equation}
in the argument of the trigonometric function.

The averaging can then be carried out analytically and the resulting transition probability~\cite{Cummings1965}
\begin{align}
\label{eq:ProbCohStateApprox}
    P_g \cong \frac{1}{2}\left[1+\exp \left(-\frac{\Theta^2}{8\,|\alpha|^2}\right)\cos{\Theta}\right]
\end{align}
shows no longer a revival due to the continuous approximation.
Yet, it still captures the initial dephasing by the Gaussian damping of the oscillation.

However, for constant $\Theta$ and increasing $|\alpha|^2$ this effect becomes less important.
Consequently, when the exponential approaches unity we recover from Eq.~\eqref{eq:ProbCohStateApprox} the Rabi limit
\begin{equation}
    P_g\cong\cos^2(\Theta/2).
\end{equation}
In this regime, highlighted for example by a $\pi$-pulse indicated on the right column of Fig.~\ref{fig:Probabilities} by the vertical line, the diffraction processes induced by a Fock state and a coherent state behave similarly.
Indeed, most atom-optical or atom-interferometric experiments~\cite{Moskowitz1983,Kazantsev1985,Kasevich1991,Adams1994} rely on diffraction from \textit{intense} classical light fields, so that the Rabi limit suffices to describe them.

\section{Mach-Zehnder atom interferometer}
\label{sec.Mach-Zehnder_atom_interferometer}
\begin{figure}
	\centering
	\includegraphics[width=1\columnwidth]{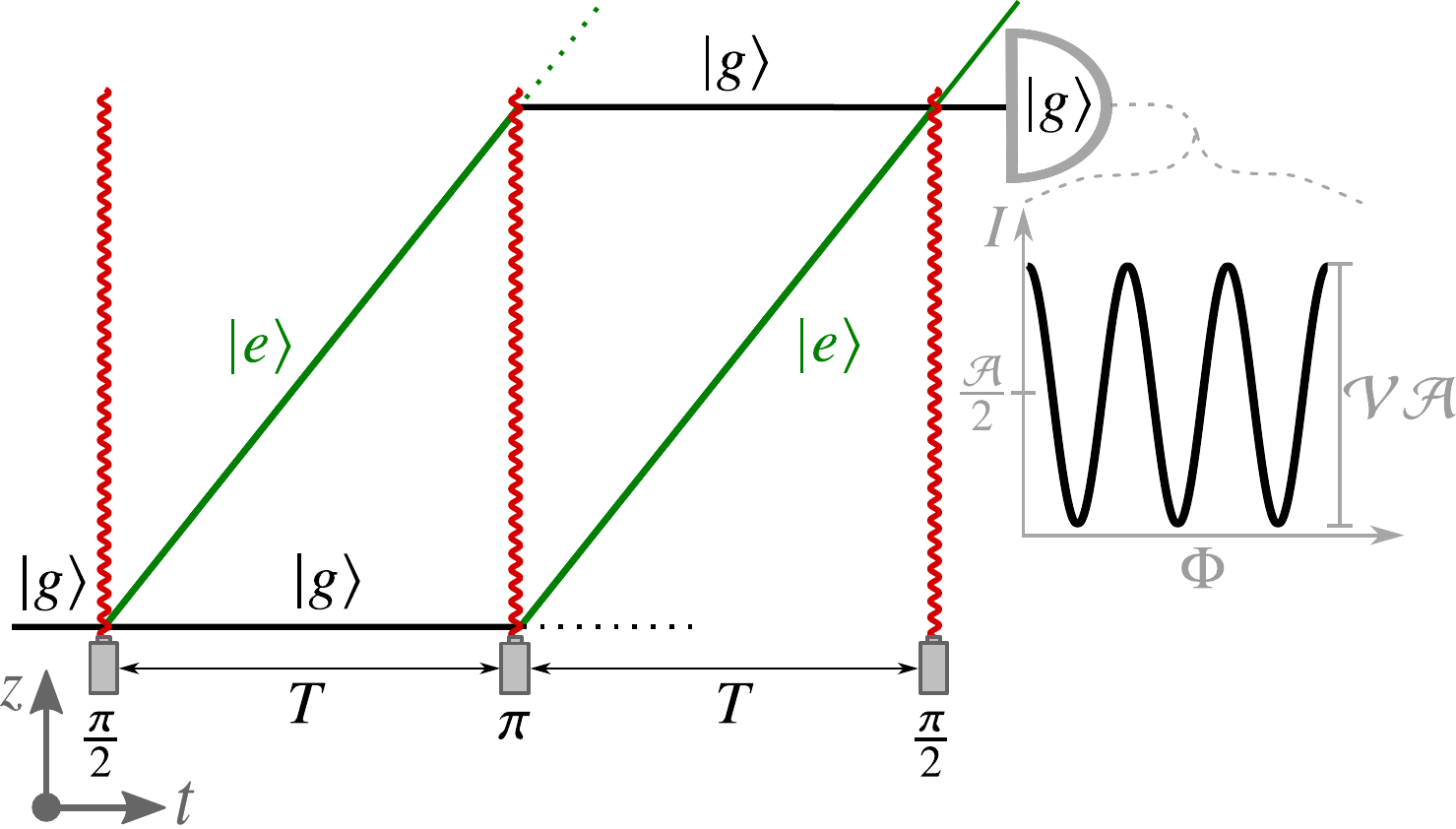}
	\caption{
	Spacetime diagram of a closed Mach-Zehnder atom interferometer in the absence of external potentials.
	The atom initially in the ground state $\ket{g}$ (black) is brought into an internal and a COM superposition by an interaction with a first light pulse that induces a $\pi/2$-pulse.
	After diffraction, the atom evolves for a time interval $T$ in a superposition of two different branches: The upper branch is associated with the excited state $\ket{e}$ (green), while the remaining ground-state population evolves along the lower branch.
	A central $\pi$-pulse redirects both branches and swaps both internal states.
	Moreover, this pulse leads in general to particle losses due to a distribution of photon numbers in the pulse area.
	Such losses are indicated by dotted lines and will not be observed by the detector.
	After the mirror pulse, both branches evolve freely for another time interval $T$.
	Finally, they are mixed by a final $\pi/2$-pulse, giving rise to an interference signal $I$.
	This signal, depicted on the right side of the figure, can be measured by postselecting on the ground state, indicated by the detector symbol.
	In this exit port we observe interference fringes with visibility $\mathcal{V}$ and amplitude $\mathcal{A}$, leading to a measurement of the phase difference $\Phi$.}
    \label{fig:MZGeometry}
\end{figure}
In Sec.~\ref{sec.Atomic_diffraction} we have set up the formalism of atom diffraction from a single quantized light field and have analyzed the associated Rabi oscillations.
We now employ these techniques to determine the interference signal of a Mach-Zehnder atom interferometer with quantum fields.

This closed light-pulse atom interferometer shown in Fig.~\ref{fig:MZGeometry} consists of a $\pi/2$-pulse (beam splitter), which creates a superposition of the COM motion and internal states.
A subsequent free propagation leads to two spatially separated branches, before a $\pi$-pulse (mirror) redirects them.
After a second free propagation, a final $\pi/2$-pulse mixes the branches and the interference signal is encoded in the ground-state population of the atom measured by the detector of Fig.~\ref{fig:MZGeometry}.

Hence, we find in one exit port the signal  
\begin{equation}
\label{eq:FullSignal}
    I= \frac{\mathcal{A}}{2}\big(1+ \mathcal{V}\cos{\Phi}\big)
\end{equation}
with amplitude $\mathcal{A}$, visibility $\mathcal{V}$ and a phase difference $\Phi$ accumulated between both branches upon propagation.

In the presence of an external gravitational field or a rotation this phase carries information about them and serves as a sensor.
However, in order to bring out most clearly the influence of the quantized fields on the interference signal we focus in this article on the interferometer in absence of external potentials.

Moreover, we neglect effects from off-resonant (detuned) diffraction or velocity selectivity~\cite{Hartman2020,Hartman2020_2}.
This treatment is justified when light shifts are compensated~\cite{McGuirk2002,Louchet2011,Friedrich2016}, and for experiments that work with sufficiently short pulses or sufficiently cold atom clouds with narrow momentum distributions, either generated by velocity selection~\cite{Kasevich1991_2} or by utilizing Bose-Einstein condensates~\cite{Torii2000}.

\subsection{Classical light fields}
A conventional Mach-Zehnder interferometer employs atom-optical elements generated by classical laser pulses.
Under the assumptions discussed above, we find $\mathcal{A}=\mathcal{V}=1$.
Because the scheme is symmetric in time as sketched in Fig.~\ref{fig:MZGeometry}, and we assume no external potential acting on the COM motion, the only surviving phase contribution is a three-point sampling 
\begin{equation}
\label{eq:ClassLightPhaseMZI}
    \Phi=\varphi_2-2\varphi_1+\varphi_0
\end{equation}
of the phases $\varphi_\ell$ for each classical laser pulse $\ell = 0, 1,$ and 2.
This \emph{laser phase contribution} is of the form of a discrete second derivative.

\subsection{Quantized light fields}

\label{sec.MZI_QLF}
We now replace the three laser pulses by quantized light fields and introduce the creation and annihilation operators $\hat{a}^{\dagger}_{\ell}$ and $\hat{a}_{\ell}$ with $\ell=0,1,$ and 2 describing the $\ell$-th pulse.
Since these operators act in different Hilbert spaces, we find the commutation relation
\begin{equation}
   \big[ \hat{a}_{\ell}^{\phantom{\dagger}},\hat{a}^{\dagger}_{\ell^\prime}\big]= \delta_{\ell,\ell^\prime}
\end{equation}
and define for each field the photon number $\hat{n}_\ell\equiv\hat{a}^{\dagger}_{\ell}\hat{a}_{\ell}$.

Since the interferometer consists of a sequence of three pulses, we add a superscript to the scattering operator of Eq.~\eqref{eq:ScattMatrix}, so that $\hat{\mathcal{S}}^{\qty(\ell)}$ describes the action of the $\ell$-th pulse on the atom.
It involves only the photonic operators $\hat{a}_\ell, \hat{a}_\ell^{\dagger}$, and $\hat{n}_\ell$.
Moreover, we denote the pulse area by $\Theta_\ell$, the mean photon number by $\bar n_\ell$, as well as the phase of the coupling constant by $\theta_\ell$.

As in the case of classical light fields, the pulse areas are chosen as $\Theta_1 = \pi$ (mirror) and $\Theta_0=\Theta_2=\pi/2$ (beam splitters).
However, the operator-valued prefactors $c_{\hat{n}_\ell}$ and $s_{\hat{n}_\ell}$ corresponding to the Rabi oscillations still depend on the photon number operator $\hat{n}_{\ell}$.
Thus, although the pulse areas are fixed, the difference between $\hat{n}_\ell$ and $\hat{n}_\ell+1$ in these prefactors is only negligible for large $\bar n_\ell$.
Note that low photon numbers would require a strong coupling because we have assumed short pulse durations.

As indicated in Fig.~\ref{fig:MZGeometry} by dotted lines, this difference in the photon number leads to a particle loss at the mirror and imperfect (unbalanced) beam splitters.
In such a situation, we expect a drop of the amplitude $\mathcal{A}$ and visibility $\mathcal{V}$ of the interference signal.

The operator sequence 
\begin{align}
\label{eq:OpSequence}
    \hat{U}_\text{MZ}\equiv \hat{\mathcal{S}}^{\qty(2)}\hat{U}\hat{\mathcal{S}}^{\qty(1)}\hat{U}\hat{\mathcal{S}}^{\qty(0)}
\end{align}
describing the unitary time evolution of the atom on its path through the Mach-Zehnder interferometer consists of the three scattering operators $\hat{\mathcal{S}}^{\qty(\ell)}$ corresponding to the three fields, as well as the free propagation
\begin{align}
    \hat{U}\equiv \exp\left[-\ii\left(\frac{\hat{p}^2}{2m\hbar}+\sum_{\ell=0}^2{ \omega \hat{n}_\ell} +\omega_a \op{e}\right)T\right]
\end{align}
for a time duration $T$ between the pulses.
Here $\hat{p}$ denotes the momentum operator of the atom of mass $m$ and all light fields have the same frequency $\omega$, whereas the energy difference between the two internal states is $\hbar \omega_a$.

For an atom initially in the ground state and detected in the same state, we postselect on the matrix element $\hat{U}^{gg}_\text{MZ}=\bra{g}\hat{U}_\text{MZ}\ket{g}$.
Moreover, since the detector is placed at a specific location, there is an additional implicit postselection of the COM motion and we can disregard the spurious paths denoted by dotted lines in  Fig.~\ref{fig:MZGeometry}.

Thus, the detection measures only the two relevant branches shown in Fig.~\ref{fig:MZGeometry} and we find the representation 
\begin{align}
    \hat{U}^{gg}_\text{MZ}=\hat{\mathcal{O}}_\text{l}+\hat{\mathcal{O}}_\text{u}
\end{align} 
where $\hat{\mathcal{O}}_\text{l}$ and $\hat{\mathcal{O}}_\text{u}$ describe the evolution along the lower and upper branch, respectively.

Hence, the interference signal 
\begin{align}
    I\equiv\big<\hat{U}^{gg\dagger}_\text{MZ}\hat{U}^{gg}_\text{MZ}\big>
\end{align} 
reads
\begin{align}
\label{eq:IntSignalOp}
    I=\big<\hat{\mathcal{O}}^{\dagger}_\text{l}\hat{\mathcal{O}}_\text{l}+\hat{\mathcal{O}}^{\dagger}_\text{u}\hat{\mathcal{O}}_\text{u}+\hat{\mathcal{O}}^{\dagger}_\text{l}\hat{\mathcal{O}}_\text{u}+\hat{\mathcal{O}}^{\dagger}_\text{u}\hat{\mathcal{O}}_\text{l}\big>.
\end{align}
Here, the first two terms determine the amplitude $\mathcal{A}/2$ and thus particle losses due to imperfect Rabi oscillations.
Conversely, the last two terms lead to the interferometer phase $\Phi$ and the visibility $\mathcal{V}$ determined by the overlap between the propagation along both branches.

Next we derive explicit expressions for $\hat{\mathcal{O}}_\text{l}$ and $\hat{\mathcal{O}}_\text{u}$.
For this purpose we note that since the Mach-Zehnder geometry without any external potential is fully symmetric, we can neglect all kinetic terms in the time evolution along each path, as they do not lead to any additional interferometer phase.
A similar argument applies to the phases that arise from the energy splitting of the atom, which oscillate with $ \omega_a$, as well as to the phases that arise from the evolution of the light fields, oscillating with $\omega$. 

Therefore, we obtain from the operator sequence of the Mach-Zehnder interferometer from Eq.~\eqref{eq:OpSequence} the operators 
\begin{subequations}
\label{eq:EffOp}
\begin{align}
    \hat{\mathcal{O}}_\text{u}&=\ee^{\ii(\theta_0-\theta_1)}c_{\hat{n}_2}\otimes \frac{s_{\hat{n}_1}}{\sqrt{\hat{n}_1}}\hat{a}^{\dagger}_1\otimes\hat{a}_0\frac{s_{\hat{n}_0}}{\sqrt{\hat{n}_0}}
\end{align}
and
\begin{align}
    \hat{\mathcal{O}}_\text{l}&=\ee^{\ii(\theta_1-\theta_2)}\frac{s_{\hat{n}_2}}{\sqrt{\hat{n}_2}}\hat{a}^{\dagger}_2\otimes\hat{a}_1\frac{s_{\hat{n}_1}}{\sqrt{\hat{n}_1}}\otimes c_{\hat{n}_0}
\end{align}
\end{subequations}
which exhibit a rather intuitive picture.

We first note that they only involve the \textit{field} operators of three modes, since the internal states and the COM motion have factored out.
Indeed, on the upper path reaching the detector in Fig.~\ref{fig:MZGeometry} marked $\ket{g}$, the atom makes a total of two transitions which occur in the first beam splitter and the mirror.
On the lower path the two transitions occur at the mirror and the second beam splitter.
Hence, the final momentum states of the two paths are identical, especially because no external potential acts on the atom.

Moreover, the sequence of field operators on the upper path corresponds to the absorption of a photon from the first beam splitter, emission at the mirror, and no transition at the second beam splitter corresponding to the operators $\hat{a}_0\,s_{\hat{n}_0}/\sqrt{\hat{n}_0}$, $s_{\hat{n}_1}/\sqrt{\hat{n}_1}\,\hat{a}^{\dagger}_1$, and $c_{\hat{n}_2}$.
The absorption and emission lead to the appearance of the difference $\theta_0-\theta_1$ of the two phase factors.

On the lower path we have no deflection at the first beam splitter, but absorption and emission at the mirror and the second beam splitter giving rise to the appearance of the terms $c_{\hat{n}_0}$, $\hat{a}_1\,s_{\hat{n}_1}/\sqrt{\hat{n}_1}$, and $s_{\hat{n}_2}/\sqrt{\hat{n}_2}\,\hat{a}^{\dagger}_2$.
Again the absorption and emission of the photon leads to the phase difference $\theta_1-\theta_2$.

When we substitute Eq.~\eqref{eq:EffOp} for $\hat{\mathcal{O}}_\text{l}$ and $\hat{\mathcal{O}}_\text{u}$ into Eq.~\eqref{eq:IntSignalOp} that describes the interference signal and recall the relation
\begin{align}
    f(\hat{n}_\ell)\hat{a}_\ell^\dagger =\hat{a}_\ell^\dagger f(\hat{n}_\ell+1), 
\end{align}
we obtain the building blocks
\begin{subequations}
\label{eq:SingleContribSignal}
\begin{align}
  &\hat{\mathcal{O}}^{\dagger}_\text{l}\hat{\mathcal{O}}_\text{u} = \ee^{\ii \, \Delta \theta}  \hat{a}_2  \frac{s_{\hat{n}_2}c_{\hat{n}_2}}{\sqrt{\hat{n}_2}} \otimes \qty( \frac{s_{\hat{n}_1}}{\sqrt{\hat{n}_1}}\hat{a}^\dagger_1)^2\otimes c_{\hat{n}_0}\hat{a}_0  \frac{s_{\hat{n}_0}}{\sqrt{\hat{n}_0}}
  \label{eq:overlap_op} \\
    &\hat{\mathcal{O}}^{\dagger}_\text{u}\hat{\mathcal{O}}_\text{u}=c^2_{\hat{n}_2}\otimes s^2_{\hat{n}_1+1}\otimes s^2_{\hat{n}_0} \\
    &\hat{\mathcal{O}}^{\dagger}_\text{l}\hat{\mathcal{O}}_\text{l}=s^2_{\hat{n}_2+1}\otimes s^2_{\hat{n}_1}\otimes c^2_{\hat{n}_0}.
\end{align}
\end{subequations}
The overlap operator $\hat{\mathcal{O}}^{\dagger}_\text{l}\hat{\mathcal{O}}_\text{u}$ includes the phase difference
\begin{equation}
    \Delta \theta \equiv \theta_2-2\theta_1+\theta_0
\end{equation}
that has the same structure as the one obtained from classical light fields.
However, $\Delta \theta$ originates from the coupling constant, which is usually assumed to be constant over the whole experiment, forcing $\Delta \theta$ to vanish.
Any additional phase difference measured by the interferometer must result from the quantum states of the light fields.

Moreover, we observe that at each beam splitter a photon is absorbed due to the appearance of the annihilation operator $\hat{a}_\ell$, whereas the mirror contributes with two-photon creations $(\hat{a}^{\dagger}_\ell)^2$.
In contrast, the contributions $\hat{\mathcal{O}}^{\dagger}_\text{u}\hat{\mathcal{O}}_\text{u}$ and $\hat{\mathcal{O}}^{\dagger}_\text{l}\hat{\mathcal{O}}_\text{l}$ leading to the amplitude only include the number operator, thus triggering particle losses and unbalanced beam splitters.

Comparing Eqs.~\eqref{eq:IntSignalOp} and \eqref{eq:SingleContribSignal} to the interference signal of Eq.~\eqref{eq:FullSignal}, we find the visibility $\mathcal{V}$ and the interferometer phase $\Phi$ from the decomposition
\begin{align}
\label{eq:VisAndPhase}
    \mathcal{V} \ee^{\ii \Phi} = \frac{2 \big<\hat{\mathcal{O}}^{\dagger}_\text{l}\hat{\mathcal{O}}_\text{u} \big>}{\big<\hat{\mathcal{O}}^{\dagger}_\text{l}\hat{\mathcal{O}}_\text{l}+\hat{\mathcal{O}}^{\dagger}_\text{u}\hat{\mathcal{O}}_\text{u}\big>}.
\end{align}
Here, we defined the amplitude $\mathcal{A}\equiv2\big<\hat{\mathcal{O}}^{\dagger}_\text{l}\hat{\mathcal{O}}_\text{l}+\hat{\mathcal{O}}^{\dagger}_\text{u}\hat{\mathcal{O}}_\text{u}\big>$.

We emphasize that we always include phases that are encoded in the states of the light fields into the phase $\Phi$, whereas $\mathcal{V}$ may take negative values if the sign of the trigonometric operators $s_{\hat{n}_\ell}$ and $c_{\hat{n}_\ell}$ flips.

\section{Examples of quantum light}
\label{sec.Examples_of_quantum_light}
We now apply the result of Sec.~\ref{sec.MZI_QLF} to calculate the interference signal for different initial states of the light fields.
In particular, we observe that the visibility obtained from coherent states approaches unity for high mean photon numbers.
Moreover, the measured phase includes the phases of the coherent states in complete analogy to classical fields.
For decreasing mean photon numbers, however, intricate effects arise from the quantized nature of the field.

We then analyze the signal of two highly non-classical states: Fock states lead to a situation where full Welcher-Weg (which-way) information is encoded into the light fields. 
As a consequence the interference vanishes.
Although this effect is less pronounced for superpositions of two Fock states, the visibility is still significantly reduced.

\subsection{Coherent state in each pulse}
We start our discussion by addressing the situation where all fields are in coherent states.
With the help of the eigenvalue equation
\begin{equation}
    \hat{a}_\ell \ket{\alpha_\ell}= |\alpha_\ell|\ee^{\ii \phi_\ell}\ket{\alpha_\ell}
\end{equation}
of a coherent state $\ket{\alpha_\ell}$ we can evaluate the expectation value $\big<\hat{\mathcal{O}}^{\dagger}_\text{l}\hat{\mathcal{O}}_\text{u} \big>$ of Eq.~\eqref{eq:overlap_op} and find
that the individual phases $\phi_\ell$ of each coherent state contribute to the interferometer phase
\begin{align}
\label{eq:phase_coh}
    \Phi= \Delta \theta + \Delta \phi
\end{align}
where 
\begin{align}
\label{eq:choerent_phases}
    \Delta \phi\equiv \phi_2-2 \phi_1 + \phi_ 0.
\end{align}
Hence, the phase $\Phi$ obtained from classical light fields corresponds to the sum of the phase difference $\Delta\theta$ of the coupling constant and the phase difference $\Delta \phi$ of the coherent states.
We emphasize that this result is independent of the mean photon number and leads to \emph{exactly} the same structure of the interferometer phase as classical laser fields.
\begin{figure}
	\centering
	\includegraphics[width=1\columnwidth]{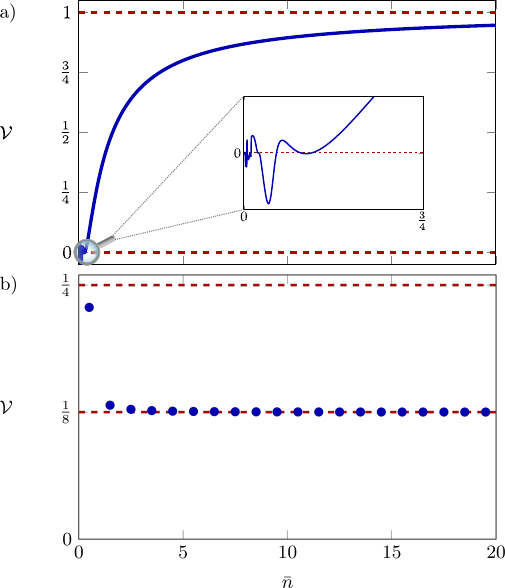}
	\caption{
	Visibility $\mathcal{V}$ of a Mach-Zehnder interferometer with all light fields in a coherent state (a) or in a superposition of two Fock states (b), as a function of the mean photon number $\bar n$:
	(a) For coherent states with $|\alpha_0|^2=|\alpha_2|^2=|\alpha_1|^2/2\equiv \bar n$ the visibility tends to unity (upper red dotted line) for large $\bar n$.
	Conversely, it drops significantly when the mean photon number decreases.
	For $\bar{n}<1$ even phase jumps occur, indicated by the oscillations around zero in the inset, and $\mathcal{V}=0$ for $\bar{n}=0$ as expected.
	(b) For light fields in a superposition of two Fock states $\mathcal{V}\rightarrow 1/8$ (lower red dotted line) for high values of $\bar{n}_0=\bar{n}_2=\bar{n}_1/2 \equiv \bar{n}$.
    Moreover, $\mathcal{V}$ is bounded by $1/4$ (upper red dotted line) which is below the classical limit of unity.}
    \label{fig:Visibility}
\end{figure}
However, the visibility is influenced by the intensities $\bar{n}_\ell=|\alpha_\ell|^2$.

In order to bring out this point most clearly we numerically calculate the amplitude
\begin{equation}
    \mathcal{A}=2\sum\limits_{\substack{n_2,n_1,\\n_0 =0}}^\infty {W_{n_2}W_{n_1}W_{n_0}\left(c^2_{n_2}s^2_{n_1+1}s^2_{n_0}+s^2_{n_2+1}s^2_{n_1}c^2_{n_0}\right)}
\end{equation}
and the visibility
\begin{align}
    \mathcal{V}=\!\sum\limits_{\substack{n_2,n_1,\\n_0 =0}}^\infty \!{ \frac{4W_{n_2}W_{n_1}W_{n_0}|\alpha_2||\alpha_1|^2|\alpha_0|s_{n_2+1}c_{n_2+1}s_{n_1+2}s_{n_1+1}s_{n_0+1}c_{n_0}}{\mathcal{A}\,\left[(n_2+1)(n_1+2)(n_1+1)(n_0+1)\right]^{1/2}}}
\end{align}
from Eq.~\eqref{eq:VisAndPhase} in the Fock basis. 

Each of the three Poissonian photon distributions $W_{n_\ell}$ from Eq.~\eqref{eq:coherent_state_distribution} has the corresponding mean value $|\alpha_\ell|^2$, and we have chosen $|\alpha_0|^2=|\alpha_2|^2=|\alpha_1|^2/2\equiv|\alpha|^2$ to account for the higher intensity of the $\pi$-pulse.
Even though the Rabi oscillations are damped as shown in Fig.~\ref{fig:Probabilities}, we find for the visibility the classical limit $\mathcal{V}\rightarrow 1$ for $|\alpha|^2\rightarrow\infty$, as shown in Fig.~\ref{fig:Visibility}(a).

This result is to be expected from the discussion of Eq.~\eqref{eq:ProbCohStateApprox}.
Indeed, if the pulse area $\Theta$ is constant but $|\alpha|^2$ increases, the damping becomes less important and is negligible in this limit, similar to the diffraction from classical fields.
In particular, the initial dephasing becomes less important for fixed pulse areas of $\pi$ or $\pi/2$.
Such transitions from quantum to classical fields have previously been studied experimentally~\cite{Bertet2001} in a Ramsey sequence consisting of two $\pi/2$-pulses and without COM motion.

However, for small mean photon numbers $|\alpha|^2$ when only a few Fock states are relevant, we observe a significant difference to the interference obtained with classical light fields.
Indeed, in this regime the visibility drops and displays small-amplitude oscillations around zero, so that even phase jumps of $\pi$ occur, as shown in the inset of Fig.~\ref{fig:Visibility}(a).
Consequently, in this limit the discrete nature of the photon distribution is relevant.

To study such effects in more detail, we analyze in the following the interference signal for Fock states and superpositions of two Fock states in each light field.

\subsection{Fock state in one pulse}
For a Fock state in one of the three light fields, we find the expectation value
\begin{align}
    \bra{n_\ell}\hat{\mathcal{O}}^{\dagger}_\text{l}\hat{\mathcal{O}}_\text{u}\ket{n_\ell} = 0
\end{align}
due to the orthogonality of the Fock basis.
Indeed, according to the expression for $ \hat{\mathcal{O}}^{\dagger}_\text{l}\hat{\mathcal{O}}_\text{u}$ from Eq.~\eqref{eq:overlap_op}, at least one creation or annihilation operator acts on every light field, leading to a vanishing visibility.

As a consequence, no phase measurement is possible as long as at least one field is in a Fock state.
Additionally, this result makes the question of the phase of a Fock state irrelevant in such an experiment.

From a more fundamental point of view, this result can be interpreted as having \textit{Welcher-Weg} information of the particle.
When measuring the photon number of the field that was initially in a Fock state, one knows whether the atom was diffracted or not, and on which branch it propagated.
As such, this result displays an illuminating example of the duality between Welcher-Weg information and interference~\cite{Agarwal2003}.

In an alternative interpretation, the COM motion of the atom is maximally entangled with the light field after the interaction with a Fock state.
Tracing out the light field therefore leads to a completely mixed state of the atom, preventing any spatial interference experiment.
In the subspace of the atom, this loss of contrast can be seen as a decoherence due to the coupling to the Fock state.

\subsection{Superposition of two Fock states in each pulse}
To avoid the encoding of \emph{complete} Welcher-Weg information, we now consider a \emph{superposition of two} Fock states in each light field.
Hence, up to a global phase factor, we write the state $\ket{\Psi}$ of the initial fields in the form 
\begin{subequations}
\begin{equation}
   \ket{\Psi} \equiv \ket{\psi_0}\otimes\ket{\psi_1}\otimes\ket{\psi_2}, 
\end{equation}
where
\begin{equation}
    \ket{\psi_\ell}\equiv\gamma_\ell \ee^{-\ii \delta_\ell/2 }  \ket{m_\ell} +\eta_\ell \ee^{\ii \delta_\ell/2 }  \ket{n_\ell}.
\end{equation}
\end{subequations}
Here, $\delta_\ell$ is the relative phase between the two Fock states, and $\gamma_\ell$ and $\eta_\ell$ are real parameters with $\gamma_\ell^2+\eta_\ell^2=1$.

To obtain a non-vanishing visibility, the structure of $\hat{\mathcal{O}}^{\dagger}_\text{l}\hat{\mathcal{O}}_\text{u}$ in Eq.~\eqref{eq:overlap_op} suggests the choice $m_0= n_0-1$, $m_1=n_1-2$, and $m_2=n_2-1$, which yields the expression
\begin{align}\label{eq.two_Fock_states}
\begin{split}
   \big<\hat{\mathcal{O}}^{\dagger}_\text{l}\hat{\mathcal{O}}_\text{u} \big> &=  \ee^{\ii \Phi }c_{n_2}\,s_{n_2}\, 
   s_{n_1-1}\,s_{n_1}\,
   c_{n_0-1}\,s_{n_0}  \prod_{\ell=0}^2 \gamma_\ell\eta_\ell,
\end{split}
\end{align}
with the interferometer phase
\begin{equation}
    \Phi = \Delta\theta+\Delta \delta
\end{equation}
where
\begin{equation}
    \Delta \delta \equiv \delta_2-\delta_1+\delta_0.
\end{equation}
Similar to the phase $\Phi$ of an interferometer with coherent states given by Eq.~\eqref{eq:phase_coh}, the coupling constant contributes with $\Delta \theta$.
However, even though the phases $\delta_\ell$ between the individual Fock states enter, the structure of $\Delta \delta$ is different from the one in $\delta \phi$ defined in Eq.~\eqref{eq:choerent_phases} for coherent states: The mirror pulse only contributes with the phase $\delta_1$, whereas the phase of a coherent state contributed with a factor of two.
Thus, no three-point sampling akin a second discrete derivative occurs, and the phase does in general not resemble the one obtained from classical light.

Equation~\eqref{eq.two_Fock_states} suggests the optimal visibility for equal superpositions $\gamma_\ell=\eta_\ell=1/\sqrt{2}$.
For this choice, we show in Fig.~\ref{fig:Visibility}(b) the visibility defined by Eq.~\eqref{eq:VisAndPhase} for different values $\bar{n}_0=\bar{n}_2=\bar{n}_1/2 \equiv \bar{n}$.
For large mean photon numbers $\bar n$ the visibility has the lower limit $\mathcal{V}\rightarrow 1/8$ and the amplitude tends to $\mathcal{A}\rightarrow 1$.
In this limit, the trigonometric functions approach classical Rabi oscillations or perfect pulses, as the difference between $\hat{n}$ and $\hat{n}+1$ becomes negligible and the visibility is solely determined by the product $\prod_{\ell=0}^2 \gamma_\ell\eta_\ell=1/8$.

However, we can optimize the individual pulses also for low photon numbers, for example through the first pulse by $c_{n_0-1}\,s_{n_0}$ yielding almost unity for $\bar n=1/2$, as depicted in Fig.~\ref{fig:Visibility}(b).
Nevertheless, the visibility has then still an upper bound of $\mathcal{V}< 1/4$ which arises from two conditions:

(i) The inequality 
\begin{equation}
    c_{n_2}\,s_{n_2}\leq 1/2
\end{equation}
arises because both trigonometric functions have \emph{identical} arguments. 

(ii) Moreover, the bound
\begin{equation}
    s_{n_1-1}\,s_{n_1}\,c_{n_0-1}\,s_{n_0}<1
\end{equation}
emerges because the trigonometric functions have \emph{different} arguments.
Together with all other prefactors in the visibility, we are therefore left with $1/4$ as an upper limit.

This result indicates that a complete information transfer from the atom to the light field is not possible with a superposition of Fock states.
Hence, we expect that for superpositions of an increasing number of Fock states the visibility increases as well, until finally reaching unity.

\section{Conclusions}
\label{sec.Conclusions}
We have determined the interference signal of a Mach-Zehnder atom interferometer taking into account the quantized nature of the light fields interacting with the atoms.
In contrast to classical fields, averaging the Rabi oscillations over the corresponding photon distributions leads to unbalanced beam splitters and mirrors causing particle loss, with implications on the interference signal.

In the case of coherent states in the three light fields the interferometer phase is identical to that for classical fields, but only for large mean photon numbers does the visibility approach the classical limit of unity.
This behavior is a consequence of the dephasing caused by the averaged Rabi oscillations, where the familiar damping becomes less dominant when the mean photon number increases for a fixed pulse area.

However, in the limit of low mean photon numbers the quantized nature of the coherent state leads to a significant drop in the visibility.
To highlight the characteristic features of this regime, we have first studied a situation where \emph{one} of the three fields is in a Fock state.
In this case the light field contains \emph{complete} Welcher-Weg information after diffraction, and no interference can be observed.
As an alternative interpretation, the light field and the COM state of the atom are maximally entangled after the diffracting process.

For a superposition of \emph{two} Fock states in \emph{all three} light fields, we regain a non-vanishing visibility $\mathcal{V}$.
However, it is bounded by $\mathcal{V}<1/4$ and the pulse areas and states can be optimized to reach this upper bound.

In addition, also the interferometer phase differs from the one obtained from coherent states:
Even superpositions of Fock states with large photon numbers are not sufficient to restore the phase structure of classical light fields. 
Conversely, the expected three-point sampling akin a second discrete derivative does in general not occur.

Our results underline that the interference observed in light-pulse atom interferometers depends on the quantum nature of the diffracting light pulses.
In contrast to diffraction patterns governed by the photon distribution, atom interferometers highlight the similarity between classical light and coherent states, although in general the interference pattern strongly depends on the quantum nature of the field.

Building on recently explored cavity schemes~\cite{Hamilton2015,Canuel2018} for atom interferometry and ideas to generate entangled COM states~\cite{Bernon2011,Haine2016,Shankar2019}, our analysis brings out additional effects of quantum light on atom interferometers.

Throughout this article we have not considered the influence of entanglement between the three quantum fields on the interference signal.
Since entanglement affects the diffraction of atoms~\cite{Khosa2004}, it must reflect itself in the interference signal.
Therefore, we will study in our future work the possibility of enhancing the visibility using entangled light fields.

\begin{acknowledgments}
We thank the QUANTUS teams from Ulm and Hanover for fruitful discussions.
The project ``Metrology with interfering Unruh-DeWitt detectors'' (MIUnD) is funded by the Carl Zeiss Foundation (Carl-Zeiss-Stiftung).
The work of IQ\textsuperscript{ST} is financially supported by the Ministry of Science, Research and Art Baden-W\"urttemberg (Ministerium f\"ur Wissenschaft, Forschung und Kunst Baden-W\"urttemberg).
The QUANTUS project is supported by the German Aerospace Center (Deutsches Zentrum f\"ur Luft- und Raumfahrt, DLR) with funds provided by the Federal Ministry of Economic Affairs and Energy (Bundesministerium f\"ur Wirtschaft und Energie, BMWi) due to an enactment of the German Bundestag under grant no. 50WM1956 (QUANTUS V).
E. G. thanks the German Research Foundation (Deutsche Forschungsgemeinschaft, DFG) for a Mercator Fellowship within CRC 1227 (DQ-mat).
W.P.S.  is grateful to  Texas A\&M University for a Faculty Fellowship at the Hagler Institute for Advanced Study at Texas A\&M University and to Texas A\&M AgriLife for the support of this work. 
\end{acknowledgments}

\section*{Data availability}
The data that support the findings of this study are available from the corresponding author upon reasonable request.

\bibliography{QuantumLightClean}

\end{document}